\documentstyle{l-aa}

\def\Msun{\ifmmode {\rm M}_\odot \else ${\rm M}_\odot$\fi}
\def\Lsun{\ifmmode {\rm L}_\odot \else ${\rm L}_\odot$\fi}
\def\VLSR{ \ifmmode {\rm V_{LSR}} \else V$_{\rm LSR}$ \fi}
\def\raw {\ifmmode\rightarrow\else$\rightarrow$\fi}             
\def\kms{\ifmmode {\,{\rm km\,s^{-1}}}                          
        \else {\hbox{$\,{\rm km\,s^{-1}}$}}\fi}
\def\as {\ifmmode {^{\scriptscriptstyle\prime\prime}}           
        \else $^{\scriptscriptstyle\prime\prime}$\fi}
\def\am {\ifmmode {^{\scriptscriptstyle\prime}}                 
        \else $^{\scriptscriptstyle\prime}$\fi}
\newdimen\sa  \newdimen\sb
\def\asc{\sa=.07em \sb=.03em
     \ifmmode $\rlap{.}$^{\scriptscriptstyle\prime\kern -\sb\prime}$\kern -\sa$
     \else \rlap{.}$^{\scriptscriptstyle\prime\kern -\sb\prime}$\kern -\sa\fi}
\def\amc{\sa=.08em \sb=.03em
     \ifmmode $\rlap{.}\kern\sa$^{\scriptscriptstyle\prime}$\kern-\sb$
     \else \rlap{.}\kern\sa$^{\scriptscriptstyle\prime}$\kern-\sb\fi}
\def\sup#1{\ifmmode {^{\rm #1}} \else $^{\rm #1}$\fi}   
\def\hms[#1 #2 #3.#4]{#1\sup{h}#2\sup{m}#3\sup{s}\llap.#4}      
\def\gms[#1 #2 #3]{#1\deg#2\am#3\as}                    
\def\deg{\ifmmode^\circ\else$^\circ$\fi}
\def\S0l {\ifmmode_{\mathord\odot} \else $_{\mathord\odot}$\fi} 
\def\cmm#1{\ifmmode {\,{\rm cm^{-#1}}}                  
        \else \hbox{$\,${\rm cm$^{\rm -#1}$}}\fi}

\begin{document}

   \thesaurus{   	
               11.01.2; Galaxies: Active
               11.19.1; Galaxies: ISM
	       11.09.01 NGC 2639;
	       11.09.01 NGC 5506;
	       11.09.01 Mrk 1;
	       11.09.01 Mrk 1210}
         
   \title{Molecular gas in H$_2$O megamaser active galaxies}

   \subtitle{}

   \author{F. Raluy\inst{1}\and P. Planesas\inst{1}\and L. Colina
\inst{2,}\thanks{On assignment from the Space Science Department of ESA}
}
           
   \offprints{F. Raluy (raluy@oan.es)}

   \institute{
	Observatorio Astron\'omico Nacional (IGN), 
	     Apartado 1143, E-28800, Alcal\'a de Henares,Madrid, Spain \\
	\null\hspace{0.2cm} (raluy@oan.es, planesas@oan.es)
\and
Space Telescope Science Institute, 3700 San Martin Drive, 
	Baltimore, MD 21218, USA (colina@stsci.edu)}
	
   \date{}
   \maketitle
   \begin{abstract}

We have searched for molecular gas towards the nucleus of four 
galaxies known to harbor a water vapor megamaser.
CO(1\raw0) emission of NGC 2639 and NGC 5506 was strong 
enough to allow us to map their inner regions. Weak emission from 
Mrk 1210 was detected and Mrk 1 was not detected at 
all. We report the tentative detection of the CO(2\raw1) line in NGC 5506. 
After this work, 12 of the 18 known galaxies harboring a 
water vapor megamaser have been observed in CO. 

The molecular gas content in the inner regions of water megamaser galaxies 
ranges from 5$\times$$10^7$ to 6$\times$$10^9~\Msun$. 
The circumnuclear molecular gas surface density also extends over nearly 
two orders of magnitude. 
The maser luminosity is correlated neither with the total amount of 
molecular gas found in the inner few kpc of these galaxies nor with global 
properties of the molecular gas such as surface density or filling 
factor; it is also independent of the infrared and optical 
luminosities. The only significant correlation we have found involves 
the maser luminosity and the low frequency radio continuum flux density. 
We conclude that the maser activity is intrinsically related to 
the energy of the active galactic nucleus whereas the intensity and even 
the presence of a water megamaser is independent of the molecular gas  
global properties such as the molecular gas content and surface density in 
the inner galactic regions.

We have also found a pos\-sible anti\-cor\-re\-lation between the 
molecular gas surface density and the rate of the megamaser variations.
A higher molecular gas abundance in the inner region could lead to higher 
maser variability because of larger nuclear flux variations due to the
more variable gas infall, and/or because of more frequent interactions of 
the pumping agent with molecular gas condensations.


   \end{abstract}	

\section{Introduction}

The first water megamaser was discovered towards the nucleus of NGC 
4945 (dos Santos \& Lepine 1979). Strong emission at 22 GHz was detected, 
with a luminosity about 100 times higher than that of W49, the 
most powerful galactic water maser. Until 
1985 four additional water megamasers were discovered, towards the
nuclei of the Circinus Galaxy, NGC 1068, NGC 4258 and 
NGC 3079. During the following decade no more extragalactic water 
megamasers were found. In 1994 a survey towards
active galactic nuclei was carried out by Braatz et al. (1994), leading to the 
discovery of five new water megamasers, doubling the number known to that date. 
Recently, Braatz et al. (1996) have reported the discovery of 6 additional 
water megamasers towards active galactic nuclei. The last ones up to now 
have been found in NGC 5793 (Hagiwara et al. 1997) and in NGC 3735 (Greenhill
et al. 1997b).

\begin{table*}[t]
\begin{center}
\caption{Adopted parameters for the observed water megamaser galaxies}
\begin{tabular}{lcccc}
\hline
Parameter & NGC 2639 & NGC 5506 & Mrk 1 & Mrk 1210 \\
\hline
\hline
$\alpha_{2000}$$^a$ & \hms[8 43 38.0] &\hms[14 13 14.9] &\hms [1 16 7.25] &
\hms[8 4 6.0 ] \\
$\delta_{2000}$$^a$ & \gms[50 12 20 3] &\gms[$-$3 12 26 7] &\gms[33 5 22 2] &
\gms[5 6 50 4] \\ 
Morphological type$^a$  & SA(r)a & SA pec sp & S & S \\
Activity$^a$  & LINER & Sy 2 &Sy 2 &Sy 2  \\	
Heliocentric velocity (\kms)$^a$  & 3236 &1816 & 4824 & 4043 \\
\VLSR (\kms) & 3235 &1825  & 4825 & 4030 \\
Distance (Mpc)$^b$ & 44 & 24 &64  &54  \\
Position angle (deg)$^c$  & 140 & 91 & --- & ---\\
Inclination (deg) & 58 & 80 & 56 & 3 \\
$D_{25}$ (arcmin)$^c$ & 1.8 & 2.8 & 0.8 & 0.8 \\ 
Linear scale (pc arcsec$^{-1}$) &209 &118 &312 & 260\\
$\log L_{\rm IR}$ (\Lsun)$^b$& 10.21  &10.35  & --- & 10.58\\
$\log L_{\rm FIR}$ (\Lsun)$^b$& 9.95  &  9.85 &  ---& 9.85    \\
\hline
\multicolumn{5}{l}{$^a$ Obtained from the NASA Extragalactic 
Database (NED)}\\
\multicolumn{5}{l}{$^b$ See more information in the text (Sect. 2)}\\
\multicolumn{5}{l}{$^c$ Obtained from the RC3 catalog (de Vaucouleurs et al.
1991)}\\
\end{tabular}

\end{center}
\end{table*}

Water megamasers, unlike the galactic and normal extragalactic masers, 
have been detected at
the nuclei of distant galaxies. Interferometric observations (Claussen 
\& Lo 1986; Greenhill et al. 1995a; Greenhill et al. 1996) have shown that
megamaser emission comes from within a region around the galactic nucleus 
whose radius is in general smaller than 1 pc.
Another important point is that all the galaxies harboring a water megamaser
present some level of activity. This
fact has led to suppose that the maser emission mechanism is identical
to galactic masers. The huge difference in the energy involved is explained 
in terms of the pumping source: the energy source of
the megamasers is the central object of the active 
nucleus, a much more powerful source than the central star powering the 
galactic masers. 

Braatz et al. (1997) have recently examined the conditions 
for detectability of water megamasers in terms of a variety of properties of 
the active galaxies, but not including the molecular gas content.
It is known (Heckman et al. 1989) that Seyfert 2 have abnormally large 
quantities of dust and gas if compared with Seyfert 1 galaxies. On the other 
hand LINER galaxies are thought to be simply extensions of Seyfert 2 galaxies 
to lower luminosities, photoionized by a weaker AGN spectrum (Osterbrock et al.
1993). Thus, the fact that no water megamasers have been found in Seyfert 1 
nuclei, while all of them are in either Seyfert 2 or LINER galaxies seems to 
indicate that a high concentration of molecular gas in the inner regions of the 
galaxies is a key parameter for the megamaser emission to be produced. 

To check this hypothesis and to find if the properties of the masers
are related to those of the molecular gas, we have carried out a study 
of the molecular gas content and its properties in several of the water 
vapor megamasers for which no previous data existed or were incomplete.
The megamasers we observed are four of those discovered by Braatz 
el al. in 1994: NGC 2639, NGC 5506, Mrk 1 and Mrk 1210. 
In our analysis we have used the available CO data for other 
water megamaser galaxies found in the literature (Heckman et al. 1989; 
Planesas et al. 1989; Sahai et al. 1990; Aalto et al. 1991; Wang 
et al. 1992 and Young et al. 1995).

\begin{figure*}[t]

	\vspace{6cm}
	\includegraphics{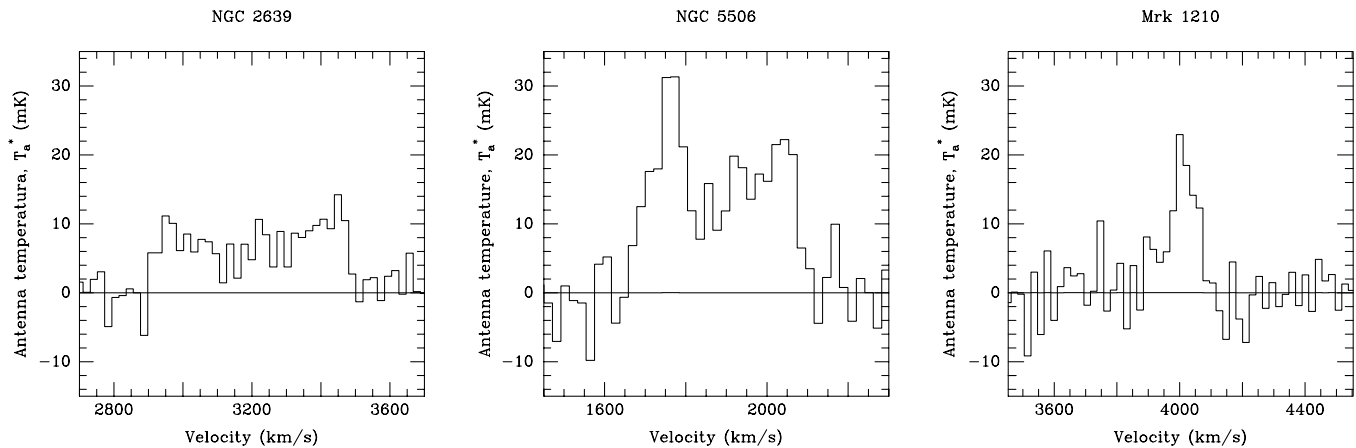}	
\caption{CO(1\raw0) emission profiles towards the central region ($23\as$ 
diameter) of the three detected water megamaser galaxies. Velocity resolution 
is $21 \kms$.
The two NGC objects show double peak spectra, which may indicate the
existence of a molecular gas ring surrounding the central AGN at a kpc
scale. The spectrum of Mrk 1210 is single-peaked, a possible indication that
the expected molecular gas ring is seen face-on. This spectrum was obtained 
in May 1997}
\end{figure*}

\section{Observations}

Most of the observations were carried out with the IRAM 30 m radiotelescope 
at Pico Veleta (Spain) in July 1995.
The transitions CO(1\raw0) at 115 GHz and CO(2\raw1) at 230 GHz were observed
simultaneously; the main beam diameters were $23\as$ and $12\as$ 
respectively, the main beam efficiencies, $\eta_b$, 0.68 and 0.40, and the 
forward efficiencies, $\eta_{fss}$, 0.92 and 0.86.  
The tentative CO(1\raw0) detection of Mrk 1210 obtained in this observing run 
prompted us to observe it again, observation that was carried out with the same 
instrument in May 1997, the antenna parameters being the same than in 1995.

A 512-channel filter-bank with a resolution of 1 MHz was used, providing a 
velocity coverage of $1200\kms$ in the CO(1\raw0) line and $600\kms$ in the 
CO(2\raw1) line, wide enough to contain the whole range of the CO emission. 
The velocity resolution was $1.3\kms$ at 230 GHz and
$2.6\kms$ at 115 GHz. The focus was checked every few hours correcting 
it when necessary. Pointing was tested every 2 hours, only small errors
were found, $2-6\as$. The typical observing time for each 
individual point was 2 hours.

All observations were made nutating the subreflector at a rate of 1 Hz, 
to a position $4\am$ in azimuth from the observed point. Thus, we were able to
obtain the flat baselines required for
the detection of the wide lines expected towards the inner 
regions of these galaxies. Linear baselines have been substracted from each 
spectrum. To obtain the absolute calibration of the
temperature scale, spectra towards IRC+10216 and Orion IRC2 at 
the redshifted frequencies were taken and compared with the IRAM 30 m catalog 
of calibrated molecular lines (Mauersberger et al. 1989).
The temperature scale in figures is the antenna temperature corrected for 
atmospheric attenuation and rear spillover, $T_{\rm A}^*$. The CO 
luminosities and derived parameters have been converted to the T$_R^*$ 
temperature to allow compatibility with the data obtained from the 
literature.

The adopted parameters for the galaxies we observed are shown in table 1.
Distances were determined from\VLSR and a 
Hubble constant of $75 \kms$~ Mpc$^{-1}$.
Infrared and far-infrared luminosities were computed
using the expressions given in Perault (1987).
Isotropic maser luminosities are given in table 2. Due to the 
strong temporal variability of the emission, they are only valid for the 
epoch its observation was carried out.

\section{Observational results}

\subsection{CO(1\raw0) line}

The spectra obtained towards the central regions of the three
detected galaxies are shown in Fig. 1. 
No emission was detected towards Mrk 1 down to a rms noise of 10 mK for a
resolution of $42 \kms$. The spectra of NGC 2639 and NGC 5506 are both
double peaked. In NGC 2639 the centroid of the CO emission is at 
$V_{\rm LSR}=3258\kms$ rather similar to the LSR velocities derived from
optical observations (3163 \kms) or from the HI 21 cm line observations 
(3298\kms).
In NGC 5506 the LSR velocity is 1827 $\kms$ from optical observations, 
$1813\kms$ from HI 21 cm line observations whereas the CO(1\raw0) centroid 
velocity is about $1880\kms$.
In Mrk 1210 the centroid velocity of the single fitted Gaussian function 
appears at $\VLSR \simeq 3995\kms$ in good agreement with values determined 
from optical observations (3968\kms) or from the HI 21 cm line observations 
(3978\kms). The intensity is 20 mK and the rms noise is 4 mK for a resolution 
of 21 \kms.

Small maps of NGC 2639 and NGC 5506 were obtained (Fig. 2). The typical 
rms noise for the thirteen NGC 2639 spectra is between 2 and 7 mK. The 
mapped region 
has a diameter of $40\as$ which corresponds to a size of 8 kpc. The 
emission is extended in the direction NW--SE, in 
agreement with the position angle of the galaxy (140$\deg$). The 
distribution of the integrated emission has an elliptical shape with 
major and minor axis of
30$\times$16$\as$ or 6$\times$3.2~kpc, in agreement with the 58$\deg$ 
inclination angle of the galaxy.

For NGC 5506, in addition to the center we observed four positions
$20\as$ away from the nucleus. The rms noise lies 
between 4 and 6 mK. 
The observed region is about 60$\as$ in diameter which corresponds 
to a size of about 7 kpc at the source distance. The emission is more 
extended towards the E--W direction, in agreement with the position angle of 
the galaxy (91$\deg$).

\begin{figure*}
\vspace{8.8cm}
\includegraphics{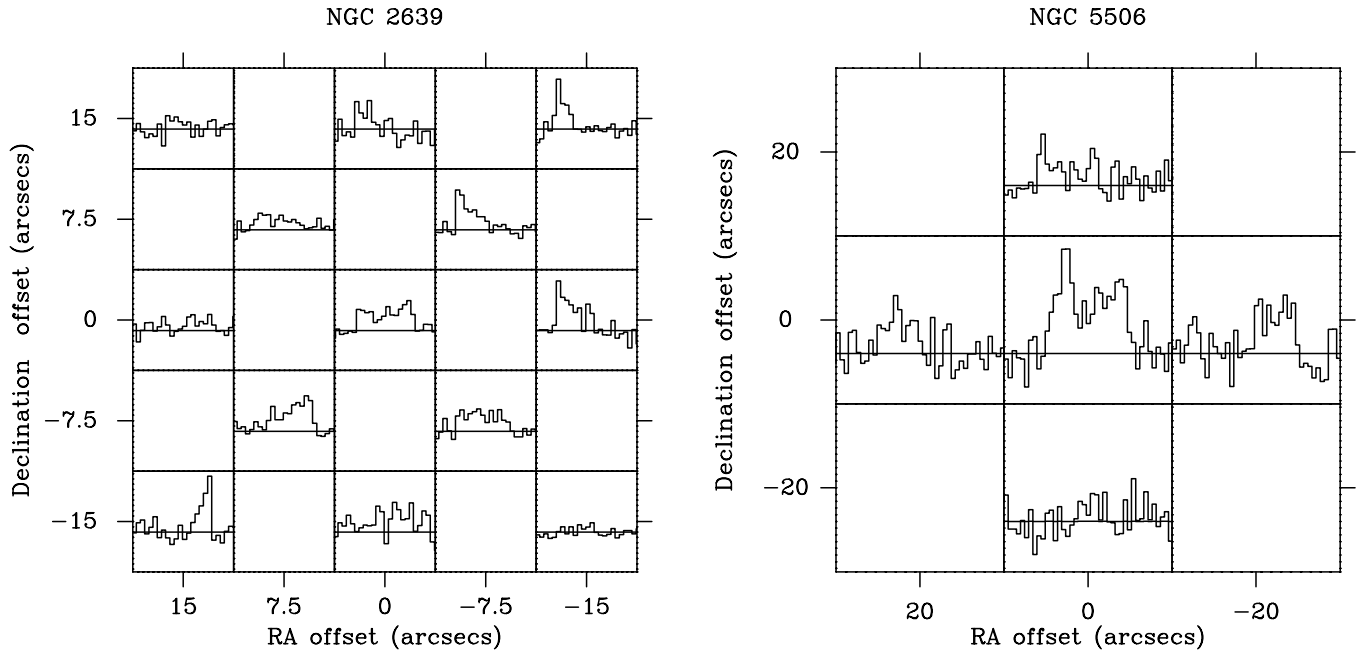}
\caption{CO(1\raw0) maps of the circumnuclear regions of NGC 2639 and NGC 5506.
In NGC 2639 the velocity range extends from 2700 to $3700 \kms$ and in 
NGC 5506 from 1450 to 2300 \kms. The resolution of the spectra is $21 \kms$ 
in NGC 5506 and $42 \kms$ in  NGC 2639}
\end{figure*}

\subsection{CO(2\raw1) line}

Observations of the CO(2\raw1) line were carried out simultaneously to those 
of the CO(1\raw0) line. We obtained a 13-point map of 
NGC 2639, a 5-point map of NGC 5506 and single central spectra in Mrk 1 
and Mrk 1210.  The rms noise achieved was 4 mK in the NGC 2639 and Mrk 1210 
spectra and 40 mK in that of Mrk 1.

We detected tentatively only the central region of NGC 5506. 
The rms noise, for a resolution of 21 \kms, is 8 mK 
and the line integrated intensity 5.5 K \kms. The centroid of the CO(2\raw1) 
emission is at 1870 \kms. In Fig. 3 (right panel) the
CO(2\raw1) spectrum produced by adding the five spectra obtained for 
this galaxy is shown. The CO(2\raw1)/CO(1\raw0) line ratio in the inner 
20\as\ region is 0.72, showing that the gas is optically thick in $^{12}$CO 
and that the excitation temperature is about 8 K (cf. Braine \& Combes, 1992). 

\begin{figure*}[t]

\vspace{6.1cm}
\includegraphics{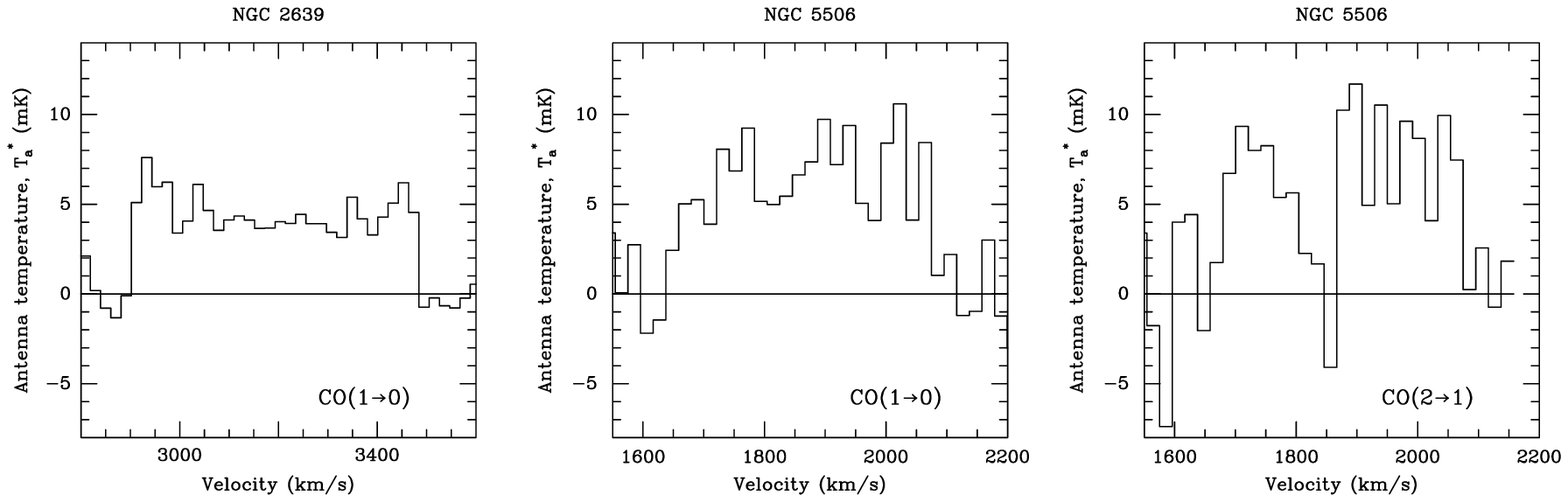}
\caption{Left and center panels: CO(1\raw0) integrated 
emission of the circumnuclear regions of NGC 2639 and NGC 5506. 
The spectra have been obtained by adding the individual spectra of all the
observed points (5 in the case of NGC 5506 and 13 in NGC 2639).
Right panel: CO(2\raw1) integrated emission of the circumnuclear region 
of NGC 5506, it has been obtained by adding the spectra of the five 
observed points.
The spectral resolution is $21\kms$ in all cases. The
intensity scale range is also the same, to allow the visual comparison of the 
intensities. 
The velocity dispersion in NGC 2639 is very high, the CO(1\raw0) line has 
a width of 600 \kms.
In NGC 5506, the emission appears whitin the same range of 
velocities (1700--2100\kms) for both observed transitions}
\end{figure*}

\subsection{CO luminosities}
The CO luminosities in table 3 have been obtained assuming a Gaussian antenna
beam, using the expression: 
$L_{\rm CO}=1.13\ d_{\rm b}^2\ I_{\rm CO}$, where $I_{\rm CO}$ 
(K($T_R^*$)\kms) is the integrated 
intensity derived from the CO(1\raw0) spectra and $d_{\rm b}$ (pc) is the 
antenna beam width at the source distance.

The H$_2$ mass has been derived from the CO luminosity using the standard 
conversion factor found in galactic giant molecular clouds, 
$M_{\rm H_2}=5.8\ L_{\rm CO}$.
High gas temperature and high metalicity could lead to an overestimate 
while high
density and photodissociation of CO could lead to an underestimate of the 
amount of molecular gas, so, we have used the standard conversion as the most
adequate choice.

The molecular gas surface density, $\Sigma_{\rm H_2}$ has been determined for 
the region of the 
galaxy included in the antenna beam, except in the galaxies where this size 
is larger (see table 2) than the CO effective diameter 
($D_{\rm CO_{\rm eff}}$, defined as the diameter that 
contains the 70 \% of the total CO emission). In the last case we used the CO 
effective diameter to calculate the surface density of molecular 
gas. To determine this size we used the optical diameter 
$D_{25}$. From the data of Kenney \& Young (1988) where CO effective diameters 
are given for a set of galaxies, we have found a correlation between 
$D_{\rm CO_{\rm eff}}$ and $D_{25}$ that is roughly independent of the model chosen for
the CO emission distribution. The relation is:
$D_{\rm CO_{\rm eff}}=0.44\ D_{25}$, very similar to that obtained by 
Young \& Scoville (1991).

In  table 2 the parameters related to the molecular gas 
are shown for the four galaxies we observed as well as for other 
water megamaser galaxies whose data have been taken from the literature.

\begin{table*}[t]
\begin{center}
\caption{Molecular gas and maser parameters in the inner region of the water 
megamaser galaxies}

\begin{tabular}{lccccccccc}
\hline
Galaxy & $\theta_{\rm B}\ ^a$ & $D_{CO_{eff}}\ ^b$ &$\log(L_{\rm CO})\ ^c$ &
$\log(M_{\rm H_2})\ ^d$ &
$L_{\rm IR}/M_{\rm H_2}$ & $\Sigma_{\rm H_2}\ ^e$ &$I_{\rm max}/I_{\rm min}$ &
$\Delta t$ & $L_{\rm maser}\ ^f$\\
 & kpc & kpc & K km s$^{-1}$ pc$^2$ & \Msun & \Lsun/\Msun
&\Msun pc$^{-2}$ & & years &\Lsun \\
\hline
\hline
NGC 2639 &4.8&10.0   &   8.1$\ ^1$   & 8.9    & 23   &21 &1.71 &1.08 &71 \\
NGC 5506 &2.7&8.8    &   7.8$\ ^1$   & 8.6    & 56   &12 & 5.43 & 1.0 &61  \\
Mrk 1    &7.2& 6.6   &$<$8.5$\ ^1$   & $<$9.3 & ---  &--- &--- &--- &64 \\
Mrk 1210  &5.7&5.6   &   7.9$\ ^1$   & 8.7    & 84   &13  &5.0 & 1.5 & 99\\
\hline
NGC 1068 &0.4& 13.0  &8.5$\ ^2$      &9.3  &75       &61 &3.& 1.25 &170 \\
NGC 3079 &3.2&14.9   &8.8$\ ^3$      &9.6  &8.4      &37 &7. & 1.25 &520 \\
NGC 4258 &1.4& 15.5  &7.4$\ ^3$      &8.1  &34       &23  & 9.0& 0.5 &85 \\
NGC 4945 &1.6&19.6   &9.1$\ ^4$      &9.8  &9.7      &270 & 1.43 & 12.5 &57\\
Circinus &1.2&5.1    &8.2$\ ^4$      &9.0  &---      &280 & 3.13 &11.7 &24\\
NGC 1386 &2.3&4.9    &7.2$\ ^5$      &8.0  &30       &6.0   & ---& --- &120 \\
NGC 5347 &8.9& 6.7   &8.3$\ ^6$      &9.1  &6.7      &18   & 1.33 &2.25 &32\\
NGC 1052 &1.6&7.6    &7.0$\ ^7$      &7.7  &38       &16  & 1.75 & 0.76 &140\\
\hline
\end{tabular}
\end{center}
$^a$ Linear size of the antenna beam at the galaxy distance\\
$^b$ Effective CO diameter, defined as the diameter that contains 70\%
of the CO emission (see more details in Sect. 3.3) \\
$^c$ CO luminosity in the central region of diameter $\theta_b$.
Obtained from the CO(1\raw0) line. Reference numbers
correspond to: (1) This paper; (2) Planesas et al. 1989; (3) Young et
al. 1995; (4) Aalto et al. 1991; (5) Sahai et al. 1990; (6) Heckman et
al. 1989; (7) Wang et al. 1992\\
$^d$ Obtained using the standard conversion factor found in
galactic GMCs, $M_{\rm H_2}= 5.8\ L_{\rm CO}$\\
$^e$ Surface density of molecular gas. It has been
obtained for a region whose diameter is the smallest value between 
$\theta_{\rm b}$, and $D_{\rm CO_{\rm eff}}$ (see details in Sect. 3.3)\\
$^f$ Isotropic maser luminosities, taken from Braatz et al. (1996)
\end{table*}

\section{Results}

\subsection{Mrk 1210}
 
Our data show that Mrk 1210 is a peculiar object among water megamaser 
galaxies. It has a relatively low surface density of molecular gas,  
not much higher than that of normal galaxies like the Milky Way or M33. 
However, its star formation efficiency, estimated from the 
$L_{\rm IR}/M_{\rm H_2}$ ratio, is the highest in our sample.
The reason may be that in this kind of objects, a significant fraction (up 
to 50 \%) of the IR luminosity could come from the active nucleus itself, 
as is the case for NGC 1068, resulting in an overestimate of the star 
formation efficiency up to a factor of two. 

Nevertheless, there are other indications of a high level of recent star 
formation activity in this galaxy.
The $s_{25}/s_{100}$ ratio (where $s_{25}$ and $s_{100}$ are the IRAS fluxes)
has been suggested as the best tracer of the ionizing photon's source 
(Dultzin-Hacyan et al. 1990). 
The high measured ratio, $s_{25}/s_{100}$=1.59, in this galaxy would thus 
indicate a large star formation activity, in 
spite of the shortage of molecular gas.

\subsection{Molecular gas properties and megamaser variability}

One of the most interesting properties of water megamasers is their 
variability. 
The observations carried out up to now have shown that the megamaser 
features vary in strength as well as in velocity (e. g. Claussen \& Lo, 1986; 
Greenhill et al. 1995b; Baan \& Haschick 1996; Braatz et al. 1996; Hagiwara 
et al. 1997). These intensity variations can reach a factor of ten.
The time scale of the variations range
between some weeks and a few years, though Greenhill et al. (1997a) have 
recently found intensity variations in the Circinus galaxy on a time
scale of a few minutes.

We have tried to determine whether long-term time variations could be related 
somehow to the abundance of molecular gas. The standard AGN model assumes that 
Seyfert 2 galaxies have a parsec-scale thick torus of gas and dust that 
blocks the direct view of the nucleus (Antonucci \& Miller 1985). 
In addition to that, a large amount of dense molecular gas has been 
found in the inner regions of these galaxies (cf. Planesas et al. 1997). 

The distribution of this gas could be relevant to the diverse phenomena 
produced by the interaction of the radiation or matter ejected from 
the nuclear source. The megamaser phenomenon could arise in the 
external layers of the nuclear torus, its long-term variability linked to the
inhomogeneities of the gas distribution in the external layers of the 
torus, but could also be affected by other structures that are known 
to exist in the inner galaxy.
In fact, a study of the circumnuclear structure of the galaxy NGC 3079 
(Baan \& Irwin 1995) showed that there are several structures of molecular gas
at different radii from the nucleus. 

On the one hand the observed megamaser variations could be related to the 
intrinsic nuclear variations. 
Intrinsic variations at different time scales in radio-continuum 
flux density have been found in one Seyfert 2 galaxy (NGC 1275, Nesterov 
et al 1995): 
slow variations have typical time scales of 20--30 years, faster 
variations having time scales ranging from some months at millimeter 
wavelengths to a few years at lower frequencies. All the radio-continuum 
variations are thought to be produced by changes in the accretion rate of 
material towards the central black hole.
In this case the molecular gas available 
near the nuclear region (scale of kpc) fuels the accretion 
disk of the AGN and any variation in the falling rate of gas towards the disk
could eventually produce the observed variation in the nuclear 
radio-continuum emission. 
The gas present at kpc scales can be transferred to the nuclear region by 
a kpc-size bar that seems to be very common in spirals (including Seyferts),
as shown by NIR observations (Mulchaey et al. 1997). 
Hence, the presence of molecular gas at a scale of hundreds 
of parsecs from the nucleus could be significant in the generation of the 
observed variability of the radio-continuum and therefore of the maser 
intensity. 

Another contribution to the observed megamaser variability could arise from
the small structure in the parsec-sized torus or in
Giant Molecular Clouds (GMCs) located not far from the nuclear source.
In particular, when one of such clouds crosses our line of sight 
towards the central maser emitting region it could act as an extra amplifier 
of the emission.
The maser variability might be due in part to the inhomogeneity of the 
molecular clouds, those regions with higher density would produce an increment 
in the observed maser intensity. Taking a maximum size of 800 AU (Falgarone et 
al. 1992) for the gas accumulations inside a cloud, and assuming 
a rotation curve similar to that of NGC 891 (e.g. Garc\'{\i}a-Burillo et al. 
1992), the time scale expected for variability produced in this way would be 
less than a few tens of 
years, compatible with the long period variability observed in megamasers. 
Accepting this reasoning, the 3 year average time-scale of the variability 
found in water megamasers (cf. Table 2) would be produced by structures ten 
times smaller.

It could be expected that in both cases, a high abundance of molecular gas
in the inner region could lead to a higher variability of the maser, via 
variable infall rate to the nucleus or via the interaction of the pumping 
source with inhomogeneous ambient clouds.
One of our results agrees with this hypothesis: we have found an 
anticorrelation (Fig. 4, center panel) between $\Sigma_{\rm H_2}$ in the 
central region of water megamaser galaxies and the rate of the relative 
variations of the megamaser intensity:
($I_{\rm max}/I_{\rm min})/\Delta t$, where $I_{\rm max}$ ($I_{\rm min}$) 
is the maximum (minimum) peak flux density observed in the megamaser 
at 22 GHz and $\Delta$t is the time elapsed between them.
The main objection that can be raised to this tentative result is the low 
spatial resolution of our determination of $\Sigma_{\rm H_2}$, as 
our measurements include a more extended region than the expected one 
relevant for the variability of the nuclear source. 

\begin{figure*}
\vspace{5.1cm}
\includegraphics{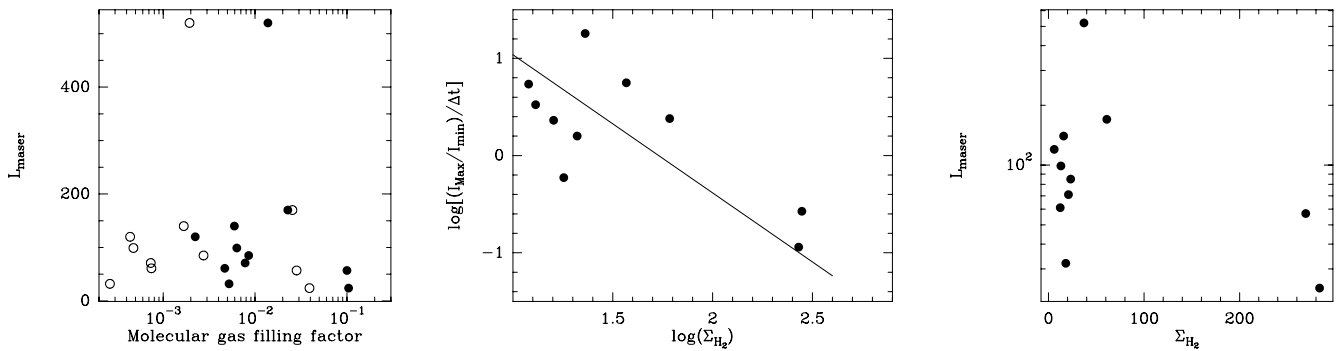}
\caption{Maser characteristics vs. molecular gas properties.
Left panel: $L_{\rm maser}$ vs. molecular gas filling factor of GMCs 
(see Sect. 4.3). To estimate the volume filling factor 
we have considered two cases: (i) a cylindrical geometry (represented in 
the figure by the black circles) with 
a diameter the size of the antenna beam at the source distance and a 
thickness of 300 pc (cf. Garc\'{\i}a-Burillo et al. 1992) and (ii) a 
spherical geometry (represented by the empty circles) with the 
diameter of the antenna beam at the source distance. 
The filling factor is a measure of the probability that a molecular cloud 
is in the line of sight of the maser. 
No correlation is found, therefore
the isotropic luminosity of the megamaser seems to be independent of the 
amount of molecular gas present in the central region of the galaxy. 
$L_{\rm maser}$ is given in \Lsun.
Center panel: $(I_{\rm max}/I_{\rm min})/\Delta t$
vs. $\Sigma_{\rm H_2}$; the first parameter is a measure of the variation 
rate of the megamaser intensity.  There is a possible correlation between the
two magnitudes (correlation coefficient of --0.71). Time is given in months 
and $\Sigma_{\rm H_2}$ in \Msun pc$^{-2}$. This result may be affected by 
the fact that the monitoring of the last discovered megamasers is still 
incomplete.
Right panel: $\log(L_{\rm maser})$ (in \Lsun) vs. $\Sigma_{\rm H_2}$
(in \Msun pc$^{-2}$). No correlation is found, see Sect. 4.3}
\end{figure*}

\subsection{Molecular gas properties and megamaser intensity}

From our estimates of $\Sigma_{\rm H_2}$ we have calculated the number 
of clouds present in the region we observed, using as the typical GMC 
parameters a diameter of 40 pc and a mass of 3$\times$$10^5 \Msun$ 
(cf. Wilson et al. 1990).
The total volume of the clouds gives us the filling factor in 
the observed region. 
Our results show that there is no correlation between the isotropic 
$L_{\rm maser}$ and the filling factor, see Fig. 4 (left panel).
This implies that the total luminosity of the megamaser is not affected 
by the abundance of molecular clouds in the inner kpc, what is reasonable
taking into account that the typical filling factor is $\sim 0.01$.

Another naively expected correlation that is not found from our data is that 
of $\log(L_{\rm maser})$ vs. 
$\Sigma_{\rm H_2}$. If we suppose that the optical depth of a molecular region
is proportional to $\Sigma_{\rm H_2}$ and that the amplification is
exponential: $I=I_0\exp{\tau}$ (Baan \& Haschick 1996), a dependence between 
$\Sigma_{\rm H_2}$ and 
$\log(L_{\rm maser})$ is expected. However, as can be seen in the Fig. 4 (right
panel) there is no correlation. Due to the complex medium the radiation 
has to go through it would not be strange that global amplification is far
from exponential. Moreover, most of the amplification is likely to take place
in a much smaller scale than the one considered when estimating 
$\Sigma_{\rm H_2}$.

\subsection{Energy of the central source and maser luminosity}

We have not found any correlation (see Fig. 5) between $L_{\rm maser}$ and 
any other relevant luminosity (i.e., $L_{\rm FIR}$, $L_{\rm IR}$, $L_{\rm B}$, 
$L_{\rm X}$). 
This is surprising, as a closer relationship between $L_{\rm FIR}$ and 
$L_{\rm X}$
with $L_{\rm maser}$ could be expected. On the one hand, the FIR radiation
is thought to be re-emission by molecular and dust clouds of the high 
energy radiation coming from the central source. On the other hand, the
adequate pumping conditions are thought to be produced by the heating of the
clouds by X-ray shocks. Therefore, although one would expect the maser 
luminosity to be related to the energy release by the central source, this 
is not the case.

A better measure of the intensity of the central source could be the 
radio continuum emission. We searched for a correlation between the maser
luminosity and the absolute low frequency radio continuum flux density of 
the galaxies. The radio continuum emission is the sum of nuclear source 
emission plus the extended emission associated with star formation; however, 
most of the radio continuum emission in active galaxies comes 
from the nucleus.
Braatz et al. (1997) have found a dependence of the 
detectability of water megamasers on intrinsic nuclear radio intensity 
at cm wavelengths. 
In the lower panels of Fig. 5 $L_{\rm maser}$  is plotted vs. the 
radio continuum flux data available for the known water megamaser galaxies.
Weak correlations are found between $L_{\rm maser}$ and the flux density at 
1.4 and 8.4 GHz, although the number of data points at 8.4 GHz is quite small.

From the correlations between $L_{\rm maser}$ and the radio continuum emission
at 1.4 and 8.4 GHz may be inferred that the maser luminosity is an 
intrinsic, energy-related property of the galactic nucleus 
(which could be characterized by the radio luminosity) 
rather than by the inner galactic regions (probably best characterized, e.g., 
by the infrared luminosities and molecular gas abundances). 

\begin{figure*}[t]
\vspace{8.4cm}
\includegraphics{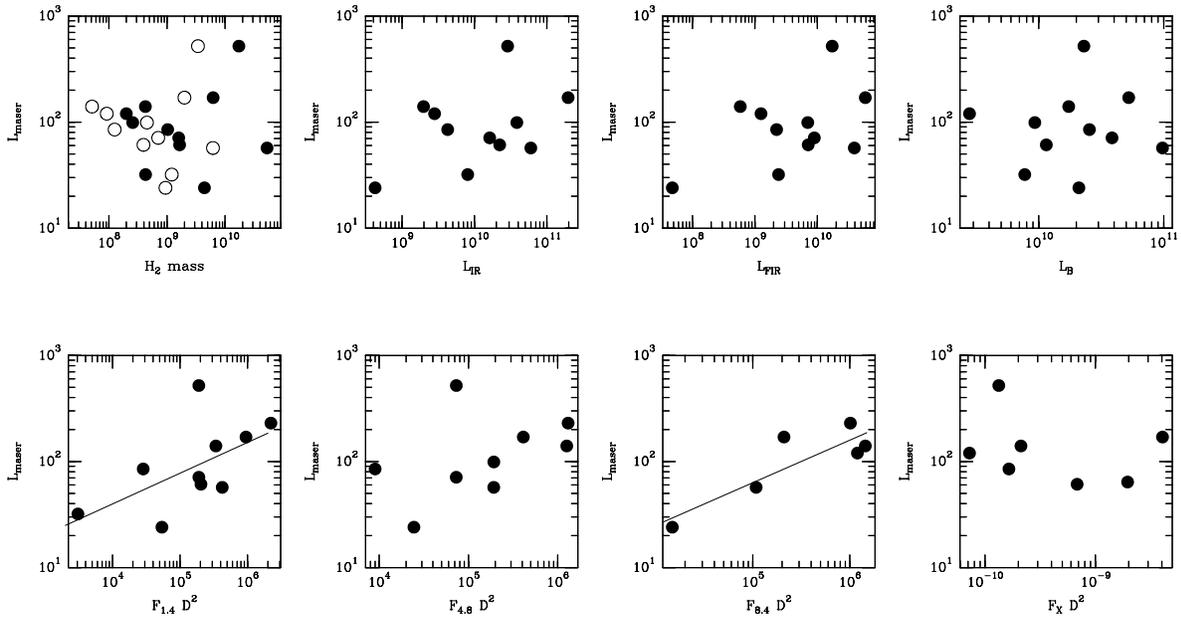}
\caption{Plot of $L_{\rm maser}$ vs. several relevant parameters 
for the water megamaser galaxies. The units of $L_{\rm IR}$, $L_{\rm FIR}$ 
and $L_{\rm maser}$ 
are \Lsun, H$_2$ mass is in \Msun. Radio fluxes are given in mJy, X-ray fluxes
in erg\ cm$^{-2}$\ s$^{-1}$ and distances in Mpc. There is a certain
correlation  between the isotropic maser luminosity and the radio continuum 
flux density at 1.4 GHz and 8.4 GHz (correlation coefficients of 0.6 and 0.87 
respectively), see Sect. 4.4} 
\end{figure*}

\section{Discussion}

We have found no correlation between the maser luminosity
and other luminosities, from far-infrared  to X-ray luminosity, 
for a sample of a dozen galaxies, in spite of the presumption that 
those luminosities could be related in one way or another to the generation 
or pumping of the maser (cf. Sect. 4.4 and Fig. 5). 
This result suggests that the maser phenomenon is very localized and 
is not related to global properties (mass or luminosity) of the galaxies. 
This is further supported by the correlation found between the low frequency 
radio continuum and the maser luminosity, both arising mostly very close to 
the galactic nucleus.

The accumulation of clouds of dense molecular gas around the 
nucleus of these galaxies seems necessary for the water megamaser
to be produced. However, we have found no apparent relation between 
the maser luminosity and 
the surface density or the total content of molecular gas in the 
inner galactic regions (say, the innermost kpc).
In spite of that, molecular gas is known to strongly shape the 
appearance of the central regions of galaxies at different
wavelengths. High angular resolution (Martin et al. 1989) and
high sensitivity CO maps (Krause et al. 1990) of 
the water megamaser galaxy NGC~4258 
have shown the importance of the molecular gas distribution in 
confining the radio jets arising from the nucleus. 
Cecil et al. (1995) have shown the coincidence of the radio jets 
and soft X-ray jets in this galaxy; 
the hot gas producing the thermal X-ray jets 
was interpreted by these authors as shocked gas pulled away 
from the ambient molecular clouds. 

The only possible correlation we have found (Fig. 4, center panel)
involves the molecular gas surface density in the inner regions
of the galaxies, in the sense that it is anticorrelated with the variation 
rate of the maser intensity.
This result can be interpreted in terms of the maser emission
being produced or further enhanced by the interaction of the nuclear 
jets with clouds of matter surrounding the active nucleus at different scales.
The combined effect of cloud movement, the small scale structure of the 
clouds and the intrinsic variability of the central source (itself related to
the gas infall towards the nucleus) could account for 
the observed variability of the megamaser intensity.
This scenario is neither meant to explain maser variability at all time scales,
nor all the subtleties observed in water maser variability, 
but to qualitatively show that one aspect of the variability 
(its rate of amplitude variations) could be somehow related to the abundance 
of molecular gas around the nucleus. 

\section{Conclusions}

We have searched for molecular gas towards the nucleus of four galaxies 
known to harbor a water vapor megamaser, and detected the CO(1\raw0) emission 
in three of them and the CO(2\raw1) emission in one. With this work 12 of the 
18 known water megamaser galaxies have been observed in CO, and only 
the most distant of the observed ones, Mrk 1, has not been detected yet.

\begin{enumerate}

	\item The 12 water megamaser galaxies with molecular gas data 
available are not an homogeneous set 
regarding their molecular gas properties. The amount of H$_2$ in their 
circumnuclear regions ranges from 5$\times$$ 10^7$ to 6$\times$$ 10^9\ \Msun$. 
The extreme values of the H$_2$ surface density, $\Sigma_{\rm H_2}$, for  
the central kpc are 6 and 280 \Msun\ pc$^{-2}$. 
This parameter extends 
over a range of 2 orders of magnitude, a range similar to that of Seyfert 
galaxies, starburst galaxies or luminous infrared galaxies. The maser 
luminosity, $L_{\rm maser}$, is not correlated to the total molecular gas mass.
Therefore it seems that the total amount of molecular gas in the inner few 
kpc is not a fundamental parameter on which depends the existence and 
the average intensity of the water megamaser.

	\item $L_{\rm maser}$ is not correlated with $\Sigma_{\rm H_2}$ or 
to the filling 
factor of giant molecular clouds. Apparently the maser luminosity does not 
depend on the 
content of molecular gas in the inner kpc. However, the accumulation of clouds 
of dense molecular gas is believed to be necessary for the generation of
a water megamaser. Observations with higher angular resolution of 
the molecular gas in the inner regions would help to solve the issue.

	\item The only correlation we have found involving the maser emission
and molecular gas parameters is between the rate of relative variation of 
the maser intensity and $\Sigma_{\rm H_2}$.
This fact may indicate that a high abundance of molecular gas in the 
inner regions could lead to higher variability in the maser emission, 
on the one hand, due to the higher variability of the central pumping source  
produced by wider variations in the gas infall; on the other hand,
due to the more frequent interactions of the pumping agent with 
molecular gas condensations.

	\item $L_{\rm maser}$ is not correlated to any other luminosity 
(infrared, optical, X-ray, blue). However, we have found 
some correlation between $L_{\rm maser}$ and the global radio continuum flux 
density at 1.4 and 8.4 GHz. This fact supports the idea that $L_{\rm maser}$ 
is a property related to the galactic nucleus (characterized by the radio
luminosities) rather than to the inner galactic 
regions (characterized by the infrared luminosity and the molecular gas 
content).

	\item Mrk 1210 stands as a peculiar object, having the 
highest star formation efficiency among water megamaser galaxies is spite 
of its relatively low molecular gas content. 

  \end{enumerate}

\begin{acknowledgements}

We thank the IRAM staff for their help during our observations at Pico Veleta.
We thank the referee, Dr. T. Wiklind for his helpful comments 
that contributed to improve the paper. 
P.P. acknowledges partial support by the Spanish DGICYT under project
PB93-0048, by the DGES project PB96-0104 and by the Collaborative Visitor 
Program at the STScI.
F.R. is fully supported by the Spanish DGICYT through the 
pre-doctoral fellowship FP94 18032957, included in the DGICYT project 
PB93-0048 and in the DGES project PB96-0104. 
This research has made use of the NASA/IPAC extragalactic database 
(NED) which is operated by the Jet Propulsion Laboratory, Caltech, 
under contract with the National Aeronautics and Space Administration.

\end{acknowledgements}


\begin{thebibliography}{}

\bibitem{}Aalto, S., Black, J.H., Johanson, L.E.B., Booth, R.S.
1991, A\&A 249, 323

\bibitem{}Antonucci, R.J., Miller, J.S. 1985, ApJ 297, 621

\bibitem{}Baan, W.A., Haschick, A. 1996, ApJ 473, 269

\bibitem{}Baan, W.A., Irwin, J.A. 1995, ApJ 446, 602

\bibitem{}Braatz, J.A., Wilson, A.S., Henkel,, C. 1994, ApJ 437, L99

\bibitem{}Braatz, J.A., Wilson, A.S., Henkel, C. 1996, ApJS 106, 51

\bibitem{}Braatz, J.A., Wilson, A.S., Henkel, C. 1997, ApJS 110, 321

\bibitem{} Braine, J., Combes, F. 1992, A\&A 264, 433

\bibitem{} Cecil, G., Wilson, A.S., de Pree, C. 1995, ApJ 440, 181

\bibitem{}Claussen, M.J., Lo, K.Y. 1986, ApJ 308, 592

\bibitem{}de Vaucouleurs, G., de Vaucouleurs, A., Corwin, H.G. et al.
1991, Third Reference Catalogue of Bright Galaxies
(Springer-Verlag: New York)

\bibitem{}dos Santos, P.M., Lepine, J.R. 1979, Nat. 278, 34

\bibitem{}Dultzin-Hacyan, D., Masegosa, J., Moles, M. 1990, A\&A 238, 28

\bibitem{}Falgarone E., Puget J.L., P\'erault M. 1993, A\&A 257, 715

\bibitem{} Garc\'{\i}a-Burillo S. Guelin, M. Cernicharo, J., Dahlem, M. 1992,
A\&A 266, 21

\bibitem{}Greenhill, L.J., Jiang, D.R., Moran, J.M. et al. 1995a,
ApJ 440, 619

\bibitem{}Greenhill, L.J., Henkel, C., Becker, T.L., Wilson, T.,
Wouterloot, J. G. A., 1995b, A\&A 304, 21

\bibitem{} Greenhill, L.J., Gwinn, C.R., Antonucci, R., Barvainis, R.
1996, ApJ 472, L21

\bibitem{}Greenhill, L.J., Ellingsen, S.P., Norris, R.P. et al.
1997a, ApJ 474, L103

\bibitem{} Greenhill, L.J., Herrnstein, J.R., Moran, J.M., Menten, K.M.,
Velusamy, T. 1997b, ApJ 486, L18 

\bibitem{} Hagiwara, Y., Kohno, K., Kawabe, R., Nakai N. 1997, PASJ 49, 171

\bibitem{}Heckman, T.M., Blitz, L., Wilson, S.A. Armus, L., Miley, G.
K. 1989, ApJ 342, 735

\bibitem{}Kenney, J., Young, J.S. 1988, ApJS 66, 261

\bibitem{} Krause, M., Cox, P., Garc\'{\i}a-Barreto, J.A., Downes, D.
1990, A\&A 233, L1

\bibitem{} Martin, P., Roy, J.R.,  Noreau, L., Lo, K.Y. 1989, ApJ 345, 707

\bibitem{} Mauersberger, R., Gu\'elin, M., Mart\'{\i}n-Pintado, J. et al.
1989, A\&AS 79, 217

\bibitem{} Mulchaey, J.S., Regan, M.W., Kundu, A. 1997, ApJS 110, 299

\bibitem{}Nesterov, N.S., Lyuty, V.M., Valtaoja, E. 1995, A\&A 296, 628

\bibitem{} Osterbrock, D.E. 1993, ApJ 404, 551

\bibitem{}Perault 1987, Th\'ese d'Etat, Universit\'e de Paris 7

\bibitem{}Planesas, P., G\'omez-Gonz\'alez, J., Mart\'{\i}n-Pintado, J., 
1989, A\&A 216, 1

\bibitem{}Planesas, P., Colina, L., P\'erez-Olea, D. 1997, A\&A 325, 81

\bibitem{}Sahai, R., Sundin, M., Claussen, M.J., Rickard, L.J. 1990,
Nordic--Baltic Astronomy Meeting

\bibitem{}Wang, Z., Kenney, J.D.P., Ishizuki, S. 1992, AJ 104, 2097

\bibitem{}Wilson, C.D., Scoville, D.  1990, ApJ 363, 435

\bibitem{}Young, J.S., Scoville, N.Z. 1991, ARA\&A 29, 581

\bibitem{}Young, J.S., Xie, J. Tacconi, L. et al.  1995, ApJS 98, 219

\end{thebibliography}
\end{document}